\documentclass[aps,prl,twocolumn]{revtex4}%
\usepackage{amsfonts}
\usepackage{amsmath}
\usepackage{amssymb}
\usepackage{graphicx}%
\providecommand{\U}[1]{\protect\rule{.1in}{.1in}}

\begin{document}
\title{Importance of subleading corrections for the Mott critical point}
\author{$^{1}$Patrick S\'{e}mon and $^{1,2}$A.-M.S. Tremblay}
\affiliation{$^{1}$D\'{e}partement de physique and Regroupement qu\'{e}b\'{e}quois sur les
mat\'{e}riaux de pointe, Universit\'{e} de Sherbrooke, Sherbrooke, Qu\'{e}bec,
Canada J1K 2R1}
\affiliation{$^{2}$Canadian Institute for Advanced Research, Toronto, Ontario, Canada, M5G 1Z8}

\begin{abstract}
The interaction-induced metal-insulator transition should be in the Ising
universality class. Experiments on layered organic superconductors suggest
that the observed critical endpoint of the first-order Mott transition belongs
instead to a different universality class. To address this question, we use
dynamical mean-field theory and a cluster generalization that is necessary to
account for short-range spatial correlations in two dimensions. Such calculations can give information
on crossover effects, in particular quantum ones, that are not included in the simplest mean-field. In the cluster calculation, a canonical transformation that minimizes the sign problem in continuous-time quantum Monte Carlo calculations allows us to obtain very
accurate results for double occupancy. These results show that there are important
subleading corrections that can lead to apparent exponents that
are different from mean-field. Experiments on optical lattices could verify our predictions.

\end{abstract}
\maketitle

Half-filled band materials should be metallic, but they are sometimes
insulators\ \cite{Imada:1998}. This paradox was discussed by Boer and Verwey
and by Peierls as early as 1937, but the first theoretical advance came from
Mott in 1949. He found that as a function of some external parameter, it is
possible to control the ratio of interaction energy to kinetic energy and
drive the system through a metal-insulator transition. This Mott transition
has by now been clearly identified in a few materials\ \cite{Imada:1998} and
in optical lattices of cold atoms\ \cite{Jordens:2008,Schneider:2008}. The
order parameter for the interaction-induced transition should be in the Ising
universality class \cite{Castellani:1979,KotliarLange:2000,Onoda:2003}, with
no breaking of translational or rotational invariance. This has been verified
explicitly in the three-dimensional compound V$_{2}$O$_{3}\;$%
\cite{Limelette:2003}$.$

It thus came as a surprise when it was discovered that in two-dimensional
layered $\kappa$-ET organic superconductors\ \cite{Lefebvre:2000}, critical exponents for
the Mott critical point, measured in both charge
(conductivity)\ \cite{Kagawa:2005} and spin (NMR) channels\ \cite{Kagawa:2009}%
, appear to belong to a different universality class. Several proposals have
appeared to explain this result. Imada \textit{et al. }%
\cite{Imada:2005,Imada:2010} suggested that while the high-temperature regime
is described by classical Ising exponents, there is also a continuous
transition at $T=0$ and, in between, a marginal quantum critical point that
controls the observed behavior. Papanikolaou \textit{et al.\ }%
\cite{papanikolaou:2008} instead started from the 2d-Ising universality class and  argued that, away from criticality, the sub-leading energy exponent dominates  for the conductivity over the leading order parameter exponent. The latter becomes relevant only very close to Tc.
A recent experiment on thermal expansion coefficient finally,
argues that the Ising universality class is the correct
one~\cite{Bartosch:2010}. That finding disagrees with the latest theoretical
calculation\ \cite{Sentef:2011} performed with Cluster Dynamical Mean-Field
theory (CDMFT)\ \cite{Kotliar:2001,Maier:2005} that measured an exponent
$\delta=2$ in agreement with the above-mentionned
conductivity\ \cite{Kagawa:2005} and NMR experiments\ \cite{Kagawa:2009}.

Here we revisit the
critical behavior at the Mott critical endpoint by studying the one-band Hubbard model,
the simplest model of interacting electrons that contains the physics of the Mott transition.
We use single-site Dynamical Mean-Field Theory
(DMFT)\ \cite{Metzner:1989,Georges:1992,Jarrell:1992} and CDMFT. Single-site
DMFT is exact in infinite dimension and can be applied to lower-dimensional
lattices\ \cite{Georges:1996,KotliarRMP:2006}. While it can be proven
analytically\ \cite{KotliarLange:2000} that for DMFT the behavior is mean-field like, the possibility of quantum critical
transients, the size of the critical region and the precise value of the
exponents have not been verified numerically. This is important since there
could be an unstable quantum critical point controlling the behavior in the
experimentally accessible regime, as has already been observed for the
conductivity\ \cite{Terletska:2011}. The same question arises with CDMFT that
takes into account the momentum dependence of the self-energy, a physical
ingredient that is known to be important in two
dimensions\ \cite{Lichtenstein:2000,Parcollet:2004,maier_d:2005,kyung:2006b,kyung:2006,Haule:2007,kancharla:2008,ohashi:2008,Sakai:2009,Liebsch:2009,Liebsch:2009b,Sordi:2010}%
. Since the Mott transition occurs far above long-ranged ordered ground
states, CDMFT should be an accurate description of the organics, except
asymptotically close to the transition where the usual critical fluctuations
should take over mean-field behavior.

Although this problem has already been studied with these methods,
improvements in computer performance and in algorithms allow us to obtain much
more accurate data. In the case of CDMFT, for the frustrated lattice considered here, the sign problem in the Continuous
Time Quantum-Monte Carlo solution of the Hybridization expansion
(CT-HYB)\ \cite{WernerMillis:2006,werner:2006,Gull:2011,Haule:2007i}
is minimized by a canonical transformation. This allows us to approach the critical point ten times closer in reduced pressure than previously possible.

\begin{figure}[ptb]
\centering{
\includegraphics[width=0.7\linewidth,clip=]{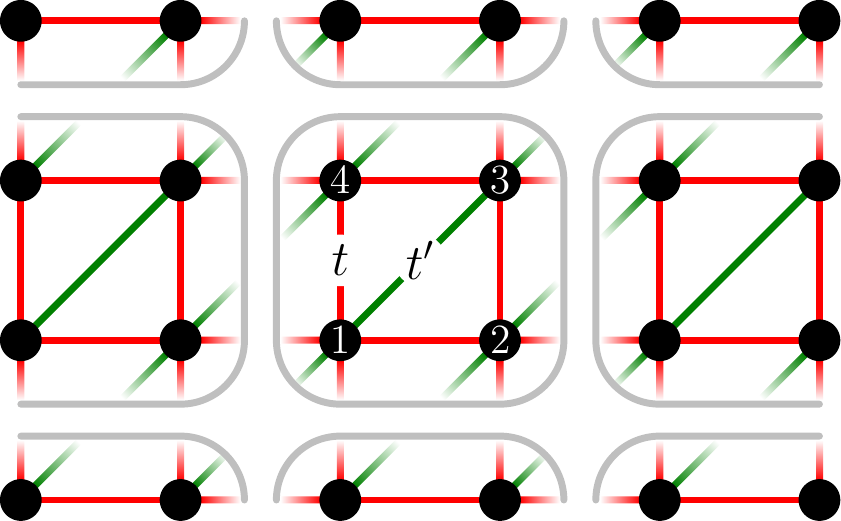}}
\caption{Periodic partitioning of the anisotropic triangular lattice into $2\times 2$ plaquettes for CDMFT.}%
\label{lattice}
\end{figure}

\textit{Method:} The simplest model that contains both the strong on-site
Coulomb repulsion and the kinetic energy of the frustrated $\kappa$-ET's
lattice, is the half-filled Hubbard model on a 2D anisotropic triangular
lattice
\begin{equation}
H=\sum_{ij\sigma}(t_{ij}-\delta_{ij}\mu)c_{i\sigma}^{\dagger}c_{j\sigma}%
+U\sum_{i}n_{i\uparrow}n_{i\downarrow},\label{Ham}%
\end{equation}
where $c_{i\sigma}^{\dagger}$ creates a spin $\sigma$ electron at site $i$,
$n_{i\sigma}=c_{i\sigma}^{\dagger}c_{i\sigma}$ is the spin $\sigma$ density at
site $i$, $t_{ij}=t_{ji}^{\ast}$ are the hopping amplitudes as shown in Fig.~(\ref{lattice}) while $\mu$ and $U$ are, respectively, the chemical potential
and the screened Coulomb repulsion.

We use single-site DMFT \cite{Georges:1996} and its cluster extension CDMFT
\cite{Kotliar:2001,Maier:2005} to solve the Hamiltonian Eq.(\ref{Ham}). These
methods start with a periodic partitioning of the infinite lattice model into
independent sites (DMFT) or clusters (CDMFT). The missing environment of the
cluster is replaced by a bath of non-interacting electrons. The action of the
cluster in a bath model may be written as
\begin{equation}
S=S_{\text{cl}}(\boldsymbol{c^{\dagger}},\boldsymbol{c})+\int_{0}^{\beta}d\tau
d\tau^{\prime}\boldsymbol{c^{\dagger}}(\tau^{\prime})\boldsymbol{\Delta}%
(\tau^{\prime}-\tau)\boldsymbol{c}(\tau),\label{Act}%
\end{equation}
where $S_{\text{cl}}$ is the cluster-action as obtained by the partitioning,
$\boldsymbol{c}$ the column vector of the corresponding $c_{i\sigma}$'s and
the bath has been integrated out in favor of a hybridization function
$\boldsymbol{\Delta}=(\Delta_{i\sigma,j\sigma^{\prime}})$. This defines an
effective impurity model. Approximating the unknown lattice self-energy
locally by the impurity self-energy $\boldsymbol{\Sigma}^{\mathcal{0}}$ and
using Dyson's equation, the lattice Green's function reads
\begin{equation}
\boldsymbol{G}_{\text{lat}}^{\mathcal{0}-1}=\boldsymbol{G}_{\text{0,lat}}%
^{-1}-\boldsymbol{\Sigma}^{\mathcal{0}},
\end{equation}
with $\boldsymbol{G}_{\text{0,lat}}$ the non-interacting lattice Green's
function. The hybridization function $\boldsymbol{\Delta}$ is determined
self-consistently from the requirement that the impurity Green's function
computed from the action Eq.\ (\ref{Act}) coincides with the projection of the
approximate lattice Green's function $\boldsymbol{G}_{\text{lat}}^{\prime}$ on
the impurity cluster. This
self-consistency condition may be derived from a variational principle
\cite{Potthoff:2003b}. For CDMFT we take the $2\times2$ plaquette illustrated
in Fig.~(\ref{lattice}). This accounts at least locally for the geometrical
frustration in the $\kappa$-ET.

To obtain the impurity Green's function (and other observables), we use a
continuous time quantum Monte-Carlo (CTQMC) solver
\cite{WernerMillis:2006,werner:2006,Gull:2011,Haule:2007i}. This method starts
with a diagrammatic expansion of the impurity partition function in powers of
the hybridization function. The key idea is then to recombine, for a given
`cluster-vertex', all possible contractions over the bath (i.e. the
hybridization function) by a determinant before summing up the series by
Monte-Carlo sampling. Without this recombination of diagrams the fermionic
sign problem would be fatal \cite{WernerMillis:2006}.

\begin{figure}[ptb]
\centering{
\includegraphics[width=1\linewidth,clip=]{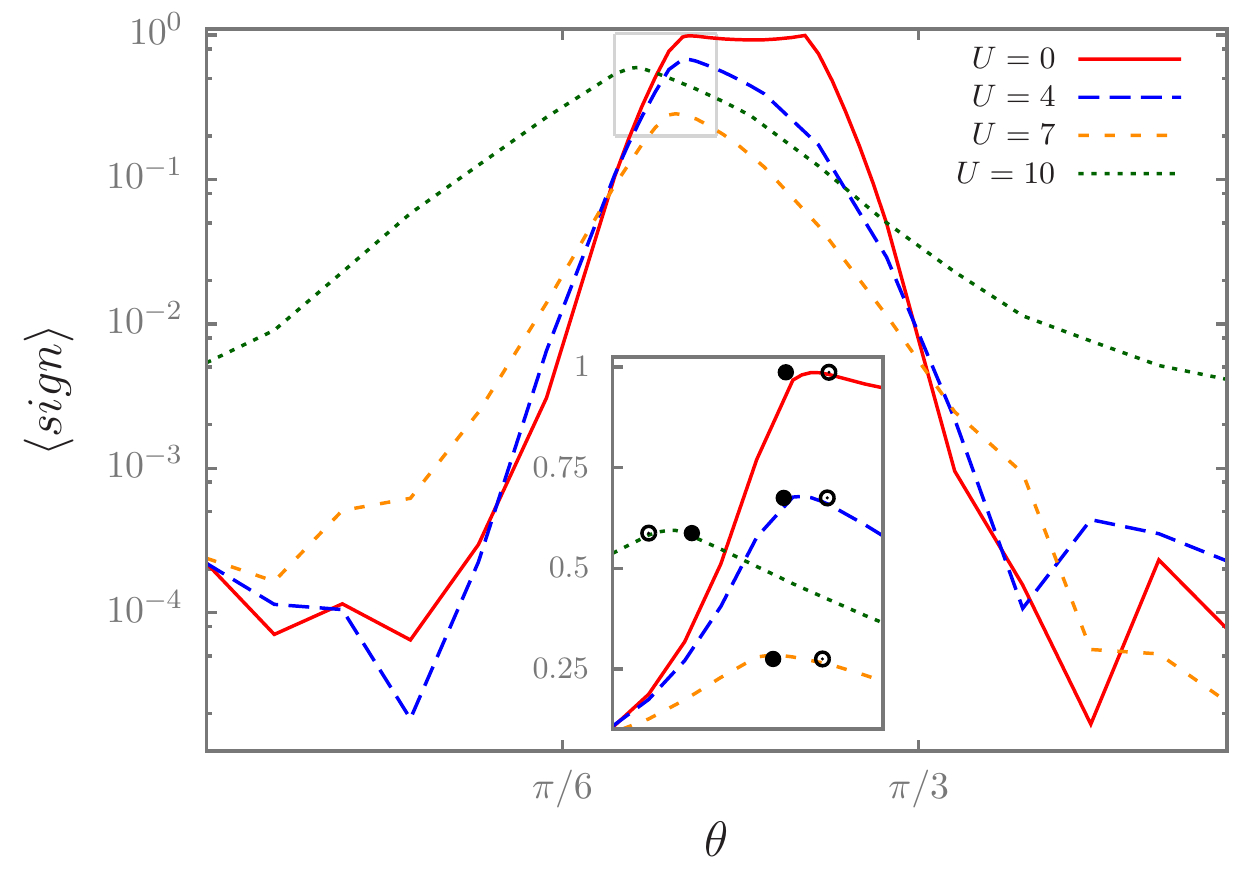}}
\caption{Average sign in CTQMC simulations of the anisotropic plaquette impurity problem at CDMFT
self-consistency with $t/t'=0.8$ ($t\equiv1$) and $\beta=20$ as a function of the angle $\theta$ in Eq.~\ref{angle}
for different values of $U$. The inset zooms on the region where the sign takes its maximum, as indicated. The dots associated with each curve indicate
the angle where the off-diagonal elements of the corresponding hybridization functions are
minimal with respect to the $L_1$ norm (filled) and the $L_2$ norm (empty).}
\label{Fig:Sign}
\end{figure}

In the case of CDMFT,
symmetries of the problem can be used to speed up the simulation
by choosing a single particle basis in Eq.~(\ref{Act}) that transforms according to the
irreducible representations~\cite{Haule:2007i}. In our case, separate charge conservation of
$\sigma=\uparrow,\downarrow$ particles and the $C_{2v}$ point
group symmetry of the anisotropic plaquette lead to the single particle basis (see Fig.~\ref{lattice} for indices)
\begin{equation}
\begin{split}
&c_{A_1\sigma} = \frac{1}{\sqrt{2}}(c_{1\sigma} + c_{3\sigma}) \quad c_{A_1\sigma}' = \frac{1}{\sqrt{2}}(c_{2\sigma} + c_{4\sigma}) \\
&c_{B_1\sigma} = \frac{1}{\sqrt{2}}(c_{1\sigma} - c_{3\sigma}) \\
&c_{B_2\sigma} = \frac{1}{\sqrt{2}}(c_{2\sigma} - c_{4\sigma}),
\end{split}
\end{equation}
with $A_1,B_1$ and $B_2$ irreducible representations of $C_{2v}$ ($A_2$ is empty).
Due to the degeneracy in the $A_1$ subspace, there is a degree of freedom in
the choice of basis which may be parameterized by an angle $\theta$ as follows,
\begin{equation}
\cos\theta c_{A_1\sigma}' - \sin\theta c_{A_1\sigma}, \quad \sin\theta c_{A_1\sigma}' + \cos\theta c_{A_1\sigma}.\label{angle}
\end{equation}
In this basis the hybridization function $\boldsymbol{\Delta}$ takes a block-diagonal
form with one $2\times 2$ block ($A_1$) and two $1\times 1$ blocks ($B_1$ and $B_2$)
for each spin (in the normal phase). The sign problem in the
Monte Carlo simulation shows a strong dependence in $\theta$ as shown in Fig.~\ref{Fig:Sign}
for $t/t'=0.8$, $\beta=20$ and different values of $U$. One can check that the maximum of the average sign is
related to the angle $\theta$ that minimizes the off-diagonal elements of the
hybridization function ($A_1$ block) with respect to some norm. The dots in the inset of Fig.~\ref{Fig:Sign}
indicate the maximum with respect to $L_1$ and $L_2$ on $[0,\beta]$. The usual basis, $\theta=0$, has a bad sign problem.

To analyze our results, we derive subleading corrections to mean-field theory. The singular part of the mean-field equation for the order parameter
$\eta$ takes the form~\cite{Kotliar:1999,KotliarLange:2000}%
\begin{equation}
p\eta+c\eta^{3}=h \label{Mean-field}%
\end{equation}
with $c$ a constant, while $p$ and $h$ are defined by $p\equiv p_{1}\left(
U-U_{c}\right)  +p_{2}\left(  T-T_{c}\right)  $ and $h\equiv h_{1}\left(
U-U_{c}\right)  +h_{2}\left(  T-T_{c}\right)  .$ Like in the liquid-gas
transition, interaction strength and temperature are not in general
eigendirections, which explains the way they appear in $p$ and $h.$ When
$p=0$, the solution is $\eta=(h/c)^{1/\delta}$, which defines $\delta=3$.
Approaching the critical line along $\delta U\equiv\left(  U-U_{c}\right)  $
for example, the mean-field Eq.(\ref{Mean-field}) takes the form
\begin{equation}
p_{1}\delta U\eta+c\eta^{3}=h_{1}\delta U.
\end{equation}
One can show that the general solution of that equation is of the form
\begin{equation}
\eta=%
{\displaystyle\sum\limits_{i=1}^{\infty}}
\delta U^{i/3}\eta_{i}.
\label{GeneralForm}
\end{equation}
with expansion coefficients $\eta_{i}$. The first term, $\delta U^{1/3},$ and
the subleading correction, $\delta U^{2/3},$ are the only terms that lead to
an infinite first derivative at the critical point. In the case of DMFT,
$\eta$ is the singular part of the hybridization function.

In the following we compute double occupancy $D\equiv\left\langle n_{\uparrow
}n_{\downarrow}\right\rangle .$ Double occupancy in general should be a smooth function of $\eta$
that can be expanded as a power series, a result that can be proven in DMFT~\cite{KotliarLange:2000}. Hence, even when $\eta$ is dominated by the leading term $\delta U^{1/\delta}$, the $\eta^{2}$ term of the power
series leads to subleading $\delta U^{2/\delta}$ corrections.

For the single-band Hubbard model, singular
behavior of $D$ implies singular behavior in both spin and charge channels~\cite{KotliarLange:2000}, as
follows from the following two sum rules on spin, $\chi_{sp},$ and charge,
$\chi_{ch},$ susceptibilities, $T\sum_{n}\int\frac{d^{2}q}{\left(
2\pi\right)  ^{2}}\chi_{sp}\left(  \mathbf{q,}\omega_{n}\right)  =n-2D$ and
$T\sum_{n}\int\frac{d^{2}q}{\left(  2\pi\right)  ^{2}}\chi_{ch}\left(
\mathbf{q,}\omega_{n}\right)  =n+2D-n^{2}$ where $\omega_{n}$ are Matsubara
frequencies and $\mathbf{q}$ wave vectors in the Brillouin zone.

Below the critical temperature, there is a first-order transition with a jump
in double occupancy that scales like $p^{\beta}$ with $\beta=1/2.$ It is very
difficult to obtain this exponent numerically because of hysteresis.
Similarly, the exponent for the susceptibility $\left(  \partial\eta/\partial
h\right)  _{p}\sim p^{-\gamma}$ with $\gamma=1$ requires numerical
differentiation and cannot be obtained accurately.

\begin{figure}[!h]
\centering{
\includegraphics[width=0.95\linewidth,clip=]{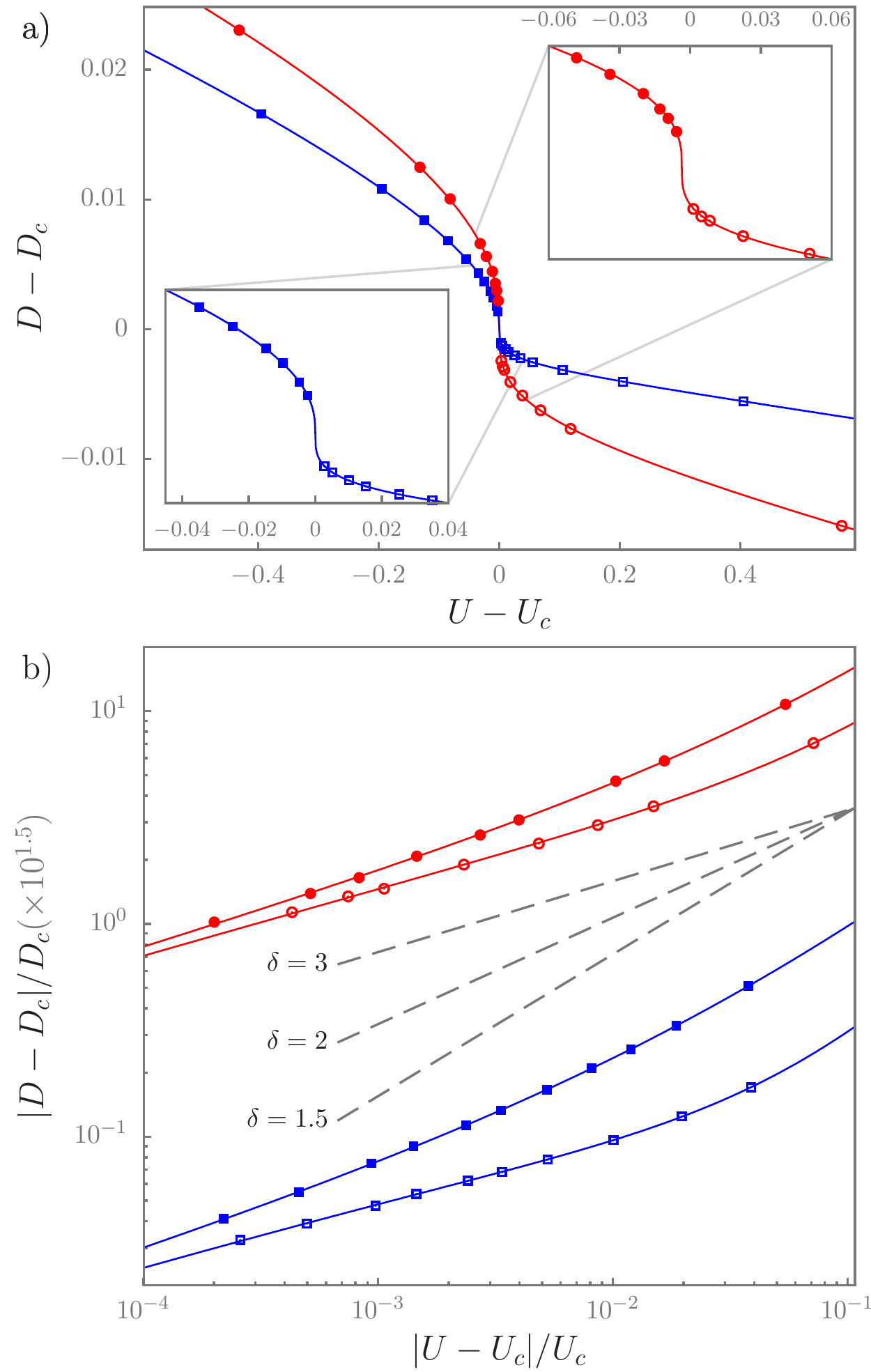}}
\caption{Double occupancy as a function of $U$ near the
Mott critical point for the Hubbard model on an anisotropic triangular lattice
with $t^{\prime}/t=0.8$ ($t\equiv1$) at half-filling and fixed critical inverse
temperature $\beta=11.15$ (squares) for DMFT and $\beta=9.9$ (circles) for CDMFT on a $2\times2$ plaquette. The solid lines show
a fit to $f(U)=c_{1}\text{sgn}(\delta U)|\delta U|^{1/\delta}+c_{2}|\delta
U|^{2/\delta}+c_{3}\delta U+D_{c}$ ($\delta U\equiv U-U_{c}$) with the same
parameters $c_{1},c_{2},c_{3},D_{c},U_{c}$ and $\delta$ for the metallic
(filled symbols) and the insulating region (open symbols). The best fitting values $(U_c, D_c, \delta)$ are $(10.445, 0.0325, 2.93)$ for DMFT and $(7.9316, 0.0679, 3.04)$ for CDMFT. (a) Linear plot centered
at $(U_{c},D_{c})$. The insets zooms on the regions close to the critical point.
(b) Logarithmic plot in reduced units relative to the critical point with
CDMFT data shifted by a factor of $10^{1.5}$ along the $y$-axis. The dashed
lines show the function $\propto|U-U_{c}|^{1/\delta}$ with $\delta$ as
indicated. In the critical regime, up to 500 iterations are necessary for convergence in the iterative
solution of the (C)DMFT equation. Once convergence is reached, we take the average over
hundreds of iterations. Monte-Carlo sweeps per iteration: $6\cdot 10^9$ for DMFT and $10^9$ for CDMFT. }%
\label{FigMott}
\end{figure}

\textit{Results:} Figure~\ref{FigMott} displays double occupancy as a function of interaction strength calculated for both single-site DMFT (blue squares) and CDMFT (red circles) at our best estimate of the corresponding critical temperatures. Both the metallic (filled symbols) and insulating (open symbols) sides are shown. The critical temperature is found as follows. Below the critical temperature, there is hysteresis and a jump in double occupancy. Above the critical temperature, double occupancy is continuous. First we searched for the highest (lowest) temperature where hysteresis (continuity) can be checked in a reasonable time. The mean of these two temperatures is then taken as an approximation for the critical temperature.

\begin{table}[!h]\centering
\small
\begin{ruledtabular}
\begin{tabular}{c c c c c c c}
\vspace{0.1cm}
&\multicolumn{2}{c}{\textbf{DFMT}}&&\multicolumn{2}{c}{\textbf{CDMFT}} \\
 \vspace{0.1cm}
 & $\delta$&$|c_2/c_1|$&& $\delta$&$|c_2/c_1|$ \\
 \hline
$D$ & 2.93  & 1.15 && 3.04 & 0.51 \\
$G_{loc}(\tau=\beta/2)$ & 2.99 & 0.32  && 3.05 & 0.33 \\
$\text{Im}\Delta_{\text{loc}}(\omega_n=\pi/\beta)$ & 3.02  & 0.28 && 3.08 & 0.086 \\
$\text{Re}\Delta_{\text{loc}}(\omega_n=\pi/\beta)$ & 2.87  & 0.79 && 3.02 & 0.75 \\
\end{tabular}
\caption{Estimates of the exponent $\delta$ from a fit of Eq.~\ref{GeneralForm} to the critical behavior of the double occupancy $D$, the local Greens function $G_{\text{loc}}$ at $\tau=\beta/2$ and the real and imaginary parts of the local hybridization $\Delta_{\text{loc}}$ function at the lowest Matsubara frequency, as obtained by DMFT and CDMFT for the same model and parameters as in Fig.~\ref{FigMott}. The ratio $|c_2/c_1|$ indicates the weight of the subleading correction, as seen from Eq~.\ref{GeneralForm}. \label{table:deltas}}
 \end{ruledtabular}
\end{table}

In Fig.~\ref{FigMott}a the scale is linear whereas it is logarithmic in Fig.~\ref{FigMott}b. The solid lines are fits to the functional form suggested by Eq.~(\ref{GeneralForm}) and by the smoothness hypothesis for $D$,
\begin{equation}
D-D_c=c_{1}\text{sgn}(\delta U)|\delta U|^{1/\delta}+c_{2}|\delta
U|^{2/\delta}+c_{3}\delta U
\label{GeneralForm}
\end{equation}
where $\delta$ and the coefficients are adjustable parameters. The fits include both the metallic and the insulating sides. We find $\delta=2.93\pm0.15$ for DMFT, where we know that the analytical result\ \cite{KotliarLange:2000} is $\delta=3$. For CDMFT we find $\delta=3.04\pm 0.25$. The errors are estimated from the values of $\delta$ at the two temperatures just below and above the critical one. The log-log plot in Fig.~\ref{FigMott}b shows that the data does not lie on a perfect straight line over the wide range of reduced units considered here. The straight dashed lines are guides to the eye that show that the exponent that we would obtain by fitting over a limited range of $\delta U$ would decrease from $\delta=3$ towards $\delta=3/2$ as we move away from the critical point. On the metallic side, the crossover extends over a rather wide region where the the exponent is close to $\delta=2$. 

As shown in Table \ref{table:deltas}, different critical quantities lead to coherent estimates of $\delta$, whereas the importance of the subleading corrections varies strongly from case to case. Note that an asymmetry of the critical quantity between the metallic and the insulating sides, as visible in the double occupancy, indicates important subleading corrections.

\textit{Discussion:} We have shown that the results of numerical
calculations with both DMFT and CDMFT are consistent with $\delta=3$ (as predicted
analytically\ \cite{KotliarLange:2000} for DMFT), if we include the subleading
corrections in the analysis. These are particularly important if the critical data is asymmetric
in the accessible region.  Otherwise, fits with a single exponent may lead to
$\delta$ closer to $\delta=2$ in the metallic phase, as observed in the
organics\ \cite{Kagawa:2005,Kagawa:2009}, and in previous CDMFT\ \cite{Sentef:2011} calculations,
where analytical results are not available. Note that in the latter calculations, the value of $t^\prime/t$ and the direction of approach to the critical point differ from ours. That may modify the size of the crossover region. 

Extracting the pressure dependence
of model parameters from band structure calculations~\cite{Kandpal:2009}, we estimate that
our numerical results are as close to the critical point in reduced units as
are the experiments. The value $\gamma=1$ in these experiments is the same as
the mean-field one, while $\beta=1$ would imply that a non-singular term
dominates the physics in the accessible range. Since coupling to the lattice
would favor mean-field exponents\ \cite{Nelson:1996}, it would be interesting
to reanalyze the experimental results by including the subleading correction
to the mean-field behavior. If the coupling to the lattice is irrelevant,
the exponents should eventually differ from mean-field behavior sufficiently
close to the transition.

To definitely settle this issue experimentally, it
would be interesting to study the two-dimensional
Mott transition in frustrated optical lattices, where double occupancy is
directly accessible\ \cite{Jordens:2008}. Our derivation of the subleading
correction also shows that it can possibly be relevant in more mundane
cases when mean-field exponents are appropriate~\cite{Brout:1960}, away from the critical regime.

\textit{Acknowledgments }We are indebted to D. S\'{e}n\'{e}chal, J. Schmalian, R. Fernandes, E. Fradkin, M. Sentef and E. Gull for
useful discussions. This work was partially supported by NSERC, the Tier I
Canada Research Chair Program (A.-M. S. T.), and Universit\'{e} de Sherbrooke.
A.-M.S.T is grateful to the Harvard Physics Department for support and P. S\'emon for hospitality during the
writing of this work. Partial support was also provided by the MIT-Harvard Center for Cold Atoms. Simulations were performed using a code based on the ALPS library~\cite{ALPS} on computers provided by CFI, MELS, Calcul Qu\'ebec and Compute Canada .
Portions of the hybridization expansion impurity solver developed by P. S\'{e}mon were inspired by the code gracefully provided by E. Gull and P. Werner.
\bigskip
\bibliographystyle{apsrev}

\begin{thebibliography}{45}
\expandafter\ifx\csname natexlab\endcsname\relax\def\natexlab#1{#1}\fi
\expandafter\ifx\csname bibnamefont\endcsname\relax
  \def\bibnamefont#1{#1}\fi
\expandafter\ifx\csname bibfnamefont\endcsname\relax
  \def\bibfnamefont#1{#1}\fi
\expandafter\ifx\csname citenamefont\endcsname\relax
  \def\citenamefont#1{#1}\fi
\expandafter\ifx\csname url\endcsname\relax
  \def\url#1{\texttt{#1}}\fi
\expandafter\ifx\csname urlprefix\endcsname\relax\def\urlprefix{URL }\fi
\providecommand{\bibinfo}[2]{#2}
\providecommand{\eprint}[2][]{\url{#2}}

\bibitem[{\citenamefont{Imada et~al.}(1998)\citenamefont{Imada, Fujimori, and
  Tokura}}]{Imada:1998}
\bibinfo{author}{\bibfnamefont{M.}~\bibnamefont{Imada}},
  \bibinfo{author}{\bibfnamefont{A.}~\bibnamefont{Fujimori}}, \bibnamefont{and}
  \bibinfo{author}{\bibfnamefont{Y.}~\bibnamefont{Tokura}},
  \bibinfo{journal}{Rev. Mod. Phys.} \textbf{\bibinfo{volume}{70}},
  \bibinfo{pages}{1039} (\bibinfo{year}{1998}),
  \urlprefix\url{http://link.aps.org/doi/10.1103/RevModPhys.70.1039}.

\bibitem[{\citenamefont{Jordens et~al.}(2008)\citenamefont{Jordens, Strohmaier,
  Gunter, Moritz, and Esslinger}}]{Jordens:2008}
\bibinfo{author}{\bibfnamefont{R.}~\bibnamefont{Jordens}},
  \bibinfo{author}{\bibfnamefont{N.}~\bibnamefont{Strohmaier}},
  \bibinfo{author}{\bibfnamefont{K.}~\bibnamefont{Gunter}},
  \bibinfo{author}{\bibfnamefont{H.}~\bibnamefont{Moritz}}, \bibnamefont{and}
  \bibinfo{author}{\bibfnamefont{T.}~\bibnamefont{Esslinger}},
  \bibinfo{journal}{Nature} \textbf{\bibinfo{volume}{455}},
  \bibinfo{pages}{204} (\bibinfo{year}{2008}), ISSN \bibinfo{issn}{0028-0836},
  \urlprefix\url{http://dx.doi.org/10.1038/nature07244}.

\bibitem[{\citenamefont{Schneider et~al.}(2008)\citenamefont{Schneider,
  Hackerm\"uller, Will, Best, Bloch, Costi, Helmes, Rasch, and
  Rosch}}]{Schneider:2008}
\bibinfo{author}{\bibfnamefont{U.}~\bibnamefont{Schneider}},
  \bibinfo{author}{\bibfnamefont{L.}~\bibnamefont{Hackerm\"uller}},
  \bibinfo{author}{\bibfnamefont{S.}~\bibnamefont{Will}},
  \bibinfo{author}{\bibfnamefont{T.}~\bibnamefont{Best}},
  \bibinfo{author}{\bibfnamefont{I.}~\bibnamefont{Bloch}},
  \bibinfo{author}{\bibfnamefont{T.~A.} \bibnamefont{Costi}},
  \bibinfo{author}{\bibfnamefont{R.~W.} \bibnamefont{Helmes}},
  \bibinfo{author}{\bibfnamefont{D.}~\bibnamefont{Rasch}}, \bibnamefont{and}
  \bibinfo{author}{\bibfnamefont{A.}~\bibnamefont{Rosch}},
  \bibinfo{journal}{Science} \textbf{\bibinfo{volume}{322}},
  \bibinfo{pages}{1520} (\bibinfo{year}{2008}),
  \eprint{http://www.sciencemag.org/content/322/5907/1520.full.pdf},
  \urlprefix\url{http://www.sciencemag.org/content/322/5907/1520.abstract}.

\bibitem[{\citenamefont{Castellani et~al.}(1979)\citenamefont{Castellani,
  Castro, Feinberg, and Ranninger}}]{Castellani:1979}
\bibinfo{author}{\bibfnamefont{C.}~\bibnamefont{Castellani}},
  \bibinfo{author}{\bibfnamefont{C.~D.} \bibnamefont{Castro}},
  \bibinfo{author}{\bibfnamefont{D.}~\bibnamefont{Feinberg}}, \bibnamefont{and}
  \bibinfo{author}{\bibfnamefont{J.}~\bibnamefont{Ranninger}},
  \bibinfo{journal}{Phys. Rev. Lett.} \textbf{\bibinfo{volume}{43}},
  \bibinfo{pages}{1957} (\bibinfo{year}{1979}),
  \urlprefix\url{http://link.aps.org/doi/10.1103/PhysRevLett.43.1957}.

\bibitem[{\citenamefont{Kotliar et~al.}(2000)\citenamefont{Kotliar, Lange, and
  Rozenberg}}]{KotliarLange:2000}
\bibinfo{author}{\bibfnamefont{G.}~\bibnamefont{Kotliar}},
  \bibinfo{author}{\bibfnamefont{E.}~\bibnamefont{Lange}}, \bibnamefont{and}
  \bibinfo{author}{\bibfnamefont{M.~J.} \bibnamefont{Rozenberg}},
  \bibinfo{journal}{Phys. Rev. Lett.} \textbf{\bibinfo{volume}{84}},
  \bibinfo{pages}{5180} (\bibinfo{year}{2000}),
  \urlprefix\url{http://link.aps.org/doi/10.1103/PhysRevLett.84.5180}.

\bibitem[{\citenamefont{Onoda and Nagaosa}(2003)}]{Onoda:2003}
\bibinfo{author}{\bibfnamefont{S.}~\bibnamefont{Onoda}} \bibnamefont{and}
  \bibinfo{author}{\bibfnamefont{N.}~\bibnamefont{Nagaosa}},
  \bibinfo{journal}{Journal of the Physical Society of Japan}
  \textbf{\bibinfo{volume}{72}}, \bibinfo{pages}{2445} (\bibinfo{year}{2003}),
  \urlprefix\url{http://jpsj.ipap.jp/link?JPSJ/72/2445/}.

\bibitem[{\citenamefont{Limelette et~al.}(2003)\citenamefont{Limelette,
  Georges, Jerome, Wzietek, Metcalf, and Honig}}]{Limelette:2003}
\bibinfo{author}{\bibfnamefont{P.}~\bibnamefont{Limelette}},
  \bibinfo{author}{\bibfnamefont{A.}~\bibnamefont{Georges}},
  \bibinfo{author}{\bibfnamefont{D.}~\bibnamefont{Jerome}},
  \bibinfo{author}{\bibfnamefont{P.}~\bibnamefont{Wzietek}},
  \bibinfo{author}{\bibfnamefont{P.}~\bibnamefont{Metcalf}}, \bibnamefont{and}
  \bibinfo{author}{\bibfnamefont{J.~M.} \bibnamefont{Honig}},
  \bibinfo{journal}{Science} \textbf{\bibinfo{volume}{302}},
  \bibinfo{pages}{89} (\bibinfo{year}{2003}),
  \eprint{http://www.sciencemag.org/cgi/reprint/302/5642/89.pdf},
  \urlprefix\url{http://www.sciencemag.org/cgi/content/abstract/302/5642/89}.

\bibitem[{\citenamefont{Lefebvre et~al.}(2000)\citenamefont{Lefebvre, Wzietek,
  Brown, Bourbonnais, J\'erome, M\'ezi\`ere, Fourmigu\'e, and
  Batail}}]{Lefebvre:2000}
\bibinfo{author}{\bibfnamefont{S.}~\bibnamefont{Lefebvre}},
  \bibinfo{author}{\bibfnamefont{P.}~\bibnamefont{Wzietek}},
  \bibinfo{author}{\bibfnamefont{S.}~\bibnamefont{Brown}},
  \bibinfo{author}{\bibfnamefont{C.}~\bibnamefont{Bourbonnais}},
  \bibinfo{author}{\bibfnamefont{D.}~\bibnamefont{J\'erome}},
  \bibinfo{author}{\bibfnamefont{C.}~\bibnamefont{M\'ezi\`ere}},
  \bibinfo{author}{\bibfnamefont{M.}~\bibnamefont{Fourmigu\'e}},
  \bibnamefont{and} \bibinfo{author}{\bibfnamefont{P.}~\bibnamefont{Batail}},
  \bibinfo{journal}{Phys. Rev. Lett.} \textbf{\bibinfo{volume}{85}},
  \bibinfo{pages}{5420} (\bibinfo{year}{2000}).

\bibitem[{\citenamefont{Kagawa et~al.}(2005)\citenamefont{Kagawa, Miyagawa, and
  Kanoda}}]{Kagawa:2005}
\bibinfo{author}{\bibfnamefont{F.}~\bibnamefont{Kagawa}},
  \bibinfo{author}{\bibfnamefont{K.}~\bibnamefont{Miyagawa}}, \bibnamefont{and}
  \bibinfo{author}{\bibfnamefont{K.}~\bibnamefont{Kanoda}},
  \bibinfo{journal}{Nature} \textbf{\bibinfo{volume}{436}},
  \bibinfo{pages}{534} (\bibinfo{year}{2005}), ISSN
  \bibinfo{issn}{{0028-0836}}.

\bibitem[{\citenamefont{Kagawa et~al.}(2009)\citenamefont{Kagawa, Miyagawa, and
  Kanoda}}]{Kagawa:2009}
\bibinfo{author}{\bibfnamefont{F.}~\bibnamefont{Kagawa}},
  \bibinfo{author}{\bibfnamefont{K.}~\bibnamefont{Miyagawa}}, \bibnamefont{and}
  \bibinfo{author}{\bibfnamefont{K.}~\bibnamefont{Kanoda}},
  \bibinfo{journal}{Nat Phys} \textbf{\bibinfo{volume}{5}},
  \bibinfo{pages}{880} (\bibinfo{year}{2009}), ISSN \bibinfo{issn}{1745-2473},
  \urlprefix\url{http://dx.doi.org/10.1038/nphys1428}.

\bibitem[{\citenamefont{Imada}(2005)}]{Imada:2005}
\bibinfo{author}{\bibfnamefont{M.}~\bibnamefont{Imada}},
  \bibinfo{journal}{Phys. Rev. B} \textbf{\bibinfo{volume}{72}},
  \bibinfo{pages}{075113} (\bibinfo{year}{2005}),
  \urlprefix\url{http://link.aps.org/doi/10.1103/PhysRevB.72.075113}.

\bibitem[{\citenamefont{Imada et~al.}(2010)\citenamefont{Imada, Misawa, and
  Yamaji}}]{Imada:2010}
\bibinfo{author}{\bibfnamefont{M.}~\bibnamefont{Imada}},
  \bibinfo{author}{\bibfnamefont{T.}~\bibnamefont{Misawa}}, \bibnamefont{and}
  \bibinfo{author}{\bibfnamefont{Y.}~\bibnamefont{Yamaji}},
  \bibinfo{journal}{Journal of Physics: Condensed Matter}
  \textbf{\bibinfo{volume}{22}}, \bibinfo{pages}{164206}
  (\bibinfo{year}{2010}),
  \urlprefix\url{http://stacks.iop.org/0953-8984/22/i=16/a=164206}.

\bibitem[{\citenamefont{Papanikolaou et~al.}(2008)\citenamefont{Papanikolaou,
  Fernandes, Fradkin, Phillips, Schmalian, and Sknepnek}}]{papanikolaou:2008}
\bibinfo{author}{\bibfnamefont{S.}~\bibnamefont{Papanikolaou}},
  \bibinfo{author}{\bibfnamefont{R.~M.} \bibnamefont{Fernandes}},
  \bibinfo{author}{\bibfnamefont{E.}~\bibnamefont{Fradkin}},
  \bibinfo{author}{\bibfnamefont{P.~W.} \bibnamefont{Phillips}},
  \bibinfo{author}{\bibfnamefont{J.}~\bibnamefont{Schmalian}},
  \bibnamefont{and} \bibinfo{author}{\bibfnamefont{R.}~\bibnamefont{Sknepnek}},
  \bibinfo{journal}{Physical Review Letters} \textbf{\bibinfo{volume}{100}},
  \bibinfo{eid}{026408} (pages~\bibinfo{numpages}{4}) (\bibinfo{year}{2008}),
  \urlprefix\url{http://link.aps.org/abstract/PRL/v100/e026408}.

\bibitem[{\citenamefont{Bartosch et~al.}(2010)\citenamefont{Bartosch, de~Souza,
  and Lang}}]{Bartosch:2010}
\bibinfo{author}{\bibfnamefont{L.}~\bibnamefont{Bartosch}},
  \bibinfo{author}{\bibfnamefont{M.}~\bibnamefont{de~Souza}}, \bibnamefont{and}
  \bibinfo{author}{\bibfnamefont{M.}~\bibnamefont{Lang}},
  \bibinfo{journal}{Phys. Rev. Lett.} \textbf{\bibinfo{volume}{104}},
  \bibinfo{pages}{245701} (\bibinfo{year}{2010}),
  \urlprefix\url{http://link.aps.org/doi/10.1103/PhysRevLett.104.245701}.

\bibitem[{\citenamefont{Sentef et~al.}()\citenamefont{Sentef, Werner, Gull, and
  Kampf}}]{Sentef:2011}
\bibinfo{author}{\bibfnamefont{M.}~\bibnamefont{Sentef}},
  \bibinfo{author}{\bibfnamefont{P.}~\bibnamefont{Werner}},
  \bibinfo{author}{\bibfnamefont{E.}~\bibnamefont{Gull}}, \bibnamefont{and}
  \bibinfo{author}{\bibfnamefont{A.~P.} \bibnamefont{Kampf}},
  \bibinfo{journal}{arXiv:1108.0428}  (????).

\bibitem[{\citenamefont{Kotliar et~al.}(2001)\citenamefont{Kotliar, Savrasov,
  P{\'a}lsson, and Biroli}}]{Kotliar:2001}
\bibinfo{author}{\bibfnamefont{G.}~\bibnamefont{Kotliar}},
  \bibinfo{author}{\bibfnamefont{S.~Y.} \bibnamefont{Savrasov}},
  \bibinfo{author}{\bibfnamefont{G.}~\bibnamefont{P{\'a}lsson}},
  \bibnamefont{and} \bibinfo{author}{\bibfnamefont{G.}~\bibnamefont{Biroli}},
  \bibinfo{journal}{Phys. Rev. Lett.} \textbf{\bibinfo{volume}{87}},
  \bibinfo{pages}{186401} (\bibinfo{year}{2001}).

\bibitem[{\citenamefont{Maier et~al.}(2005{\natexlab{a}})\citenamefont{Maier,
  Jarrell, Pruschke, and Hettler}}]{Maier:2005}
\bibinfo{author}{\bibfnamefont{T.}~\bibnamefont{Maier}},
  \bibinfo{author}{\bibfnamefont{M.}~\bibnamefont{Jarrell}},
  \bibinfo{author}{\bibfnamefont{T.}~\bibnamefont{Pruschke}}, \bibnamefont{and}
  \bibinfo{author}{\bibfnamefont{M.~H.} \bibnamefont{Hettler}},
  \bibinfo{journal}{Reviews of Modern Physics} \textbf{\bibinfo{volume}{77}},
  \bibinfo{pages}{1027} (\bibinfo{year}{2005}{\natexlab{a}}).

\bibitem[{\citenamefont{Metzner and Vollhardt}(1989)}]{Metzner:1989}
\bibinfo{author}{\bibfnamefont{W.}~\bibnamefont{Metzner}} \bibnamefont{and}
  \bibinfo{author}{\bibfnamefont{D.}~\bibnamefont{Vollhardt}},
  \bibinfo{journal}{Phys. Rev. Lett.} \textbf{\bibinfo{volume}{62}},
  \bibinfo{pages}{324} (\bibinfo{year}{1989}).

\bibitem[{\citenamefont{Georges and Kotliar}(1992)}]{Georges:1992}
\bibinfo{author}{\bibfnamefont{A.}~\bibnamefont{Georges}} \bibnamefont{and}
  \bibinfo{author}{\bibfnamefont{G.}~\bibnamefont{Kotliar}},
  \bibinfo{journal}{Phys. Rev. B} \textbf{\bibinfo{volume}{45}},
  \bibinfo{pages}{6479} (\bibinfo{year}{1992}).

\bibitem[{\citenamefont{Jarrell}(1992)}]{Jarrell:1992}
\bibinfo{author}{\bibfnamefont{M.}~\bibnamefont{Jarrell}},
  \bibinfo{journal}{Phys. Rev. Lett.} \textbf{\bibinfo{volume}{69}},
  \bibinfo{pages}{168} (\bibinfo{year}{1992}).

\bibitem[{\citenamefont{Georges et~al.}(1996)\citenamefont{Georges, Kotliar,
  Krauth, and Rozenberg}}]{Georges:1996}
\bibinfo{author}{\bibfnamefont{A.}~\bibnamefont{Georges}},
  \bibinfo{author}{\bibfnamefont{G.}~\bibnamefont{Kotliar}},
  \bibinfo{author}{\bibfnamefont{W.}~\bibnamefont{Krauth}}, \bibnamefont{and}
  \bibinfo{author}{\bibfnamefont{M.~J.} \bibnamefont{Rozenberg}},
  \bibinfo{journal}{Rev. Mod. Phys.} \textbf{\bibinfo{volume}{68}},
  \bibinfo{pages}{13 } (\bibinfo{year}{1996}).

\bibitem[{\citenamefont{Kotliar et~al.}(2006)\citenamefont{Kotliar, Savrasov,
  Haule, Oudovenko, Parcollet, and Marianetti}}]{KotliarRMP:2006}
\bibinfo{author}{\bibfnamefont{G.}~\bibnamefont{Kotliar}},
  \bibinfo{author}{\bibfnamefont{S.~Y.} \bibnamefont{Savrasov}},
  \bibinfo{author}{\bibfnamefont{K.}~\bibnamefont{Haule}},
  \bibinfo{author}{\bibfnamefont{V.~S.} \bibnamefont{Oudovenko}},
  \bibinfo{author}{\bibfnamefont{O.}~\bibnamefont{Parcollet}},
  \bibnamefont{and} \bibinfo{author}{\bibfnamefont{C.~A.}
  \bibnamefont{Marianetti}}, \bibinfo{journal}{Reviews of Modern Physics}
  \textbf{\bibinfo{volume}{78}}, \bibinfo{eid}{865}
  (pages~\bibinfo{numpages}{87}) (\bibinfo{year}{2006}),
  \urlprefix\url{http://link.aps.org/abstract/RMP/v78/p865}.

\bibitem[{\citenamefont{Terletska et~al.}(2011)\citenamefont{Terletska,
  Vu\ifmmode \check{c}\else \v{c}\fi{}i\ifmmode \check{c}\else
  \v{c}\fi{}evi\ifmmode~\acute{c}\else \'{c}\fi{},
  Tanaskovi\ifmmode~\acute{c}\else \'{c}\fi{}, and
  Dobrosavljevi\ifmmode~\acute{c}\else \'{c}\fi{}}}]{Terletska:2011}
\bibinfo{author}{\bibfnamefont{H.}~\bibnamefont{Terletska}},
  \bibinfo{author}{\bibfnamefont{J.}~\bibnamefont{Vu\ifmmode \check{c}\else
  \v{c}\fi{}i\ifmmode \check{c}\else \v{c}\fi{}evi\ifmmode~\acute{c}\else
  \'{c}\fi{}}},
  \bibinfo{author}{\bibfnamefont{D.}~\bibnamefont{Tanaskovi\ifmmode~\acute{c}\else
  \'{c}\fi{}}}, \bibnamefont{and}
  \bibinfo{author}{\bibfnamefont{V.}~\bibnamefont{Dobrosavljevi\ifmmode~\acute{c}\else
  \'{c}\fi{}}}, \bibinfo{journal}{Phys. Rev. Lett.}
  \textbf{\bibinfo{volume}{107}}, \bibinfo{pages}{026401}
  (\bibinfo{year}{2011}),
  \urlprefix\url{http://link.aps.org/doi/10.1103/PhysRevLett.107.026401}.

\bibitem[{\citenamefont{Lichtenstein and Katsnelson}(2000)}]{Lichtenstein:2000}
\bibinfo{author}{\bibfnamefont{A.~I.} \bibnamefont{Lichtenstein}}
  \bibnamefont{and} \bibinfo{author}{\bibfnamefont{M.~I.}
  \bibnamefont{Katsnelson}}, \bibinfo{journal}{Phys. Rev. B}
  \textbf{\bibinfo{volume}{62}}, \bibinfo{pages}{R9283} (\bibinfo{year}{2000}).

\bibitem[{\citenamefont{Parcollet et~al.}(2004)\citenamefont{Parcollet, Biroli,
  and Kotliar}}]{Parcollet:2004}
\bibinfo{author}{\bibfnamefont{O.}~\bibnamefont{Parcollet}},
  \bibinfo{author}{\bibfnamefont{G.}~\bibnamefont{Biroli}}, \bibnamefont{and}
  \bibinfo{author}{\bibfnamefont{G.}~\bibnamefont{Kotliar}},
  \bibinfo{journal}{Phys. Rev. Lett.} \textbf{\bibinfo{volume}{92}},
  \bibinfo{pages}{226402} (\bibinfo{year}{2004}),
  \urlprefix\url{http://link.aps.org/doi/10.1103/PhysRevLett.92.226402}.

\bibitem[{\citenamefont{Maier et~al.}(2005{\natexlab{b}})\citenamefont{Maier,
  Jarrell, Schulthess, Kent, and White}}]{maier_d:2005}
\bibinfo{author}{\bibfnamefont{T.~A.} \bibnamefont{Maier}},
  \bibinfo{author}{\bibfnamefont{M.}~\bibnamefont{Jarrell}},
  \bibinfo{author}{\bibfnamefont{T.~C.} \bibnamefont{Schulthess}},
  \bibinfo{author}{\bibfnamefont{P.~R.~C.} \bibnamefont{Kent}},
  \bibnamefont{and} \bibinfo{author}{\bibfnamefont{J.~B.} \bibnamefont{White}},
  \bibinfo{journal}{Physical Review Letters} \textbf{\bibinfo{volume}{95}},
  \bibinfo{eid}{237001} (pages~\bibinfo{numpages}{4})
  (\bibinfo{year}{2005}{\natexlab{b}}),
  \urlprefix\url{http://link.aps.org/abstract/PRL/v95/e237001}.

\bibitem[{\citenamefont{Kyung et~al.}(2006)\citenamefont{Kyung, Kancharla,
  S\'{e}n\'{e}chal, Tremblay, Civelli, and Kotliar}}]{kyung:2006b}
\bibinfo{author}{\bibfnamefont{B.}~\bibnamefont{Kyung}},
  \bibinfo{author}{\bibfnamefont{S.~S.} \bibnamefont{Kancharla}},
  \bibinfo{author}{\bibfnamefont{D.}~\bibnamefont{S\'{e}n\'{e}chal}},
  \bibinfo{author}{\bibfnamefont{A.-M.~S.} \bibnamefont{Tremblay}},
  \bibinfo{author}{\bibfnamefont{M.}~\bibnamefont{Civelli}}, \bibnamefont{and}
  \bibinfo{author}{\bibfnamefont{G.}~\bibnamefont{Kotliar}},
  \bibinfo{journal}{Physical Review B (Condensed Matter and Materials Physics)}
  \textbf{\bibinfo{volume}{73}}, \bibinfo{eid}{165114}
  (pages~\bibinfo{numpages}{6}) (\bibinfo{year}{2006}),
  \urlprefix\url{http://link.aps.org/abstract/PRB/v73/e165114}.

\bibitem[{\citenamefont{Kyung and Tremblay}(2006)}]{kyung:2006}
\bibinfo{author}{\bibfnamefont{B.}~\bibnamefont{Kyung}} \bibnamefont{and}
  \bibinfo{author}{\bibfnamefont{A.-M.~S.} \bibnamefont{Tremblay}},
  \bibinfo{journal}{Physical Review Letters} \textbf{\bibinfo{volume}{97}},
  \bibinfo{eid}{046402} (pages~\bibinfo{numpages}{4}) (\bibinfo{year}{2006}),
  \urlprefix\url{http://link.aps.org/abstract/PRL/v97/e046402}.

\bibitem[{\citenamefont{Haule and Kotliar}(2007)}]{Haule:2007}
\bibinfo{author}{\bibfnamefont{K.}~\bibnamefont{Haule}} \bibnamefont{and}
  \bibinfo{author}{\bibfnamefont{G.}~\bibnamefont{Kotliar}},
  \bibinfo{journal}{Physical Review B (Condensed Matter and Materials Physics)}
  \textbf{\bibinfo{volume}{76}}, \bibinfo{eid}{104509}
  (pages~\bibinfo{numpages}{37}) (\bibinfo{year}{2007}),
  \urlprefix\url{http://link.aps.org/abstract/PRB/v76/e104509}.

\bibitem[{\citenamefont{Kancharla et~al.}(2008)\citenamefont{Kancharla, Kyung,
  Senechal, Civelli, Capone, Kotliar, and Tremblay}}]{kancharla:2008}
\bibinfo{author}{\bibfnamefont{S.~S.} \bibnamefont{Kancharla}},
  \bibinfo{author}{\bibfnamefont{B.}~\bibnamefont{Kyung}},
  \bibinfo{author}{\bibfnamefont{D.}~\bibnamefont{Senechal}},
  \bibinfo{author}{\bibfnamefont{M.}~\bibnamefont{Civelli}},
  \bibinfo{author}{\bibfnamefont{M.}~\bibnamefont{Capone}},
  \bibinfo{author}{\bibfnamefont{G.}~\bibnamefont{Kotliar}}, \bibnamefont{and}
  \bibinfo{author}{\bibfnamefont{A.-M.~S.} \bibnamefont{Tremblay}},
  \bibinfo{journal}{Physical Review B (Condensed Matter and Materials Physics)}
  \textbf{\bibinfo{volume}{77}}, \bibinfo{eid}{184516}
  (pages~\bibinfo{numpages}{12}) (\bibinfo{year}{2008}),
  \urlprefix\url{http://link.aps.org/abstract/PRB/v77/e184516}.

\bibitem[{\citenamefont{Ohashi et~al.}(2008)\citenamefont{Ohashi, Momoi,
  Tsunetsugu, and Kawakami}}]{ohashi:2008}
\bibinfo{author}{\bibfnamefont{T.}~\bibnamefont{Ohashi}},
  \bibinfo{author}{\bibfnamefont{T.}~\bibnamefont{Momoi}},
  \bibinfo{author}{\bibfnamefont{H.}~\bibnamefont{Tsunetsugu}},
  \bibnamefont{and} \bibinfo{author}{\bibfnamefont{N.}~\bibnamefont{Kawakami}},
  \bibinfo{journal}{Physical Review Letters} \textbf{\bibinfo{volume}{100}},
  \bibinfo{eid}{076402} (pages~\bibinfo{numpages}{4}) (\bibinfo{year}{2008}),
  \urlprefix\url{http://link.aps.org/abstract/PRL/v100/e076402}.

\bibitem[{\citenamefont{Sakai et~al.}(2009)\citenamefont{Sakai, Motome, and
  Imada}}]{Sakai:2009}
\bibinfo{author}{\bibfnamefont{S.}~\bibnamefont{Sakai}},
  \bibinfo{author}{\bibfnamefont{Y.}~\bibnamefont{Motome}}, \bibnamefont{and}
  \bibinfo{author}{\bibfnamefont{M.}~\bibnamefont{Imada}},
  \bibinfo{journal}{Phys. Rev. Lett.} \textbf{\bibinfo{volume}{102}},
  \bibinfo{pages}{056404} (\bibinfo{year}{2009}),
  \urlprefix\url{http://link.aps.org/doi/10.1103/PhysRevLett.102.056404}.

\bibitem[{\citenamefont{Liebsch and Tong}(2009)}]{Liebsch:2009}
\bibinfo{author}{\bibfnamefont{A.}~\bibnamefont{Liebsch}} \bibnamefont{and}
  \bibinfo{author}{\bibfnamefont{N.-H.} \bibnamefont{Tong}},
  \bibinfo{journal}{Phys. Rev. B} \textbf{\bibinfo{volume}{80}},
  \bibinfo{pages}{165126} (\bibinfo{year}{2009}),
  \urlprefix\url{http://link.aps.org/doi/10.1103/PhysRevB.80.165126}.

\bibitem[{\citenamefont{Liebsch et~al.}(2009)\citenamefont{Liebsch, Ishida, and
  Merino}}]{Liebsch:2009b}
\bibinfo{author}{\bibfnamefont{A.}~\bibnamefont{Liebsch}},
  \bibinfo{author}{\bibfnamefont{H.}~\bibnamefont{Ishida}}, \bibnamefont{and}
  \bibinfo{author}{\bibfnamefont{J.}~\bibnamefont{Merino}},
  \bibinfo{journal}{Phys. Rev. B} \textbf{\bibinfo{volume}{79}},
  \bibinfo{pages}{195108} (\bibinfo{year}{2009}),
  \urlprefix\url{http://link.aps.org/doi/10.1103/PhysRevB.79.195108}.

\bibitem[{\citenamefont{Sordi et~al.}(2010)\citenamefont{Sordi, Haule, and
  Tremblay}}]{Sordi:2010}
\bibinfo{author}{\bibfnamefont{G.}~\bibnamefont{Sordi}},
  \bibinfo{author}{\bibfnamefont{K.}~\bibnamefont{Haule}}, \bibnamefont{and}
  \bibinfo{author}{\bibfnamefont{A.~M.~S.} \bibnamefont{Tremblay}},
  \bibinfo{journal}{Phys. Rev. Lett.} \textbf{\bibinfo{volume}{104}},
  \bibinfo{pages}{226402} (\bibinfo{year}{2010}).

\bibitem[{\citenamefont{Werner et~al.}(2006)\citenamefont{Werner, Comanac,
  de\char39{} Medici, Troyer, and Millis}}]{WernerMillis:2006}
\bibinfo{author}{\bibfnamefont{P.}~\bibnamefont{Werner}},
  \bibinfo{author}{\bibfnamefont{A.}~\bibnamefont{Comanac}},
  \bibinfo{author}{\bibfnamefont{L.}~\bibnamefont{de\char39{} Medici}},
  \bibinfo{author}{\bibfnamefont{M.}~\bibnamefont{Troyer}}, \bibnamefont{and}
  \bibinfo{author}{\bibfnamefont{A.~J.} \bibnamefont{Millis}},
  \bibinfo{journal}{Phys. Rev. Lett.} \textbf{\bibinfo{volume}{97}},
  \bibinfo{pages}{076405} (\bibinfo{year}{2006}).

\bibitem[{\citenamefont{Werner and Millis}(2006)}]{werner:2006}
\bibinfo{author}{\bibfnamefont{P.}~\bibnamefont{Werner}} \bibnamefont{and}
  \bibinfo{author}{\bibfnamefont{A.~J.} \bibnamefont{Millis}},
  \bibinfo{journal}{Physical Review B (Condensed Matter and Materials Physics)}
  \textbf{\bibinfo{volume}{74}}, \bibinfo{eid}{155107}
  (pages~\bibinfo{numpages}{13}) (\bibinfo{year}{2006}),
  \urlprefix\url{http://link.aps.org/abstract/PRB/v74/e155107}.

\bibitem[{\citenamefont{Gull et~al.}(2011)\citenamefont{Gull, Millis,
  Lichtenstein, Rubtsov, Troyer, and Werner}}]{Gull:2011}
\bibinfo{author}{\bibfnamefont{E.}~\bibnamefont{Gull}},
  \bibinfo{author}{\bibfnamefont{A.~J.} \bibnamefont{Millis}},
  \bibinfo{author}{\bibfnamefont{A.~I.} \bibnamefont{Lichtenstein}},
  \bibinfo{author}{\bibfnamefont{A.~N.} \bibnamefont{Rubtsov}},
  \bibinfo{author}{\bibfnamefont{M.}~\bibnamefont{Troyer}}, \bibnamefont{and}
  \bibinfo{author}{\bibfnamefont{P.}~\bibnamefont{Werner}},
  \bibinfo{journal}{Rev. Mod. Phys.} \textbf{\bibinfo{volume}{83}},
  \bibinfo{pages}{349} (\bibinfo{year}{2011}),
  \urlprefix\url{http://link.aps.org/doi/10.1103/RevModPhys.83.349}.

\bibitem[{\citenamefont{Haule}(2007)}]{Haule:2007i}
\bibinfo{author}{\bibfnamefont{K.}~\bibnamefont{Haule}},
  \bibinfo{journal}{Physical Review B (Condensed Matter and Materials Physics)}
  \textbf{\bibinfo{volume}{75}}, \bibinfo{eid}{155113}
  (pages~\bibinfo{numpages}{12}) (\bibinfo{year}{2007}).

\bibitem[{\citenamefont{Potthoff}(2003)}]{Potthoff:2003b}
\bibinfo{author}{\bibfnamefont{M.}~\bibnamefont{Potthoff}},
  \bibinfo{journal}{Eur. Phys. J. B (France)} \textbf{\bibinfo{volume}{32}},
  \bibinfo{pages}{429 } (\bibinfo{year}{2003}).

\bibitem[{\citenamefont{Kotliar}(1999)}]{Kotliar:1999}
\bibinfo{author}{\bibfnamefont{G.}~\bibnamefont{Kotliar}},
  \bibinfo{journal}{The European Physical Journal B - Condensed Matter and
  Complex Systems} \textbf{\bibinfo{volume}{11}}, \bibinfo{pages}{27}
  (\bibinfo{year}{1999}), ISSN \bibinfo{issn}{1434-6028},
  \bibinfo{note}{10.1007/s100510050914},
  \urlprefix\url{http://dx.doi.org/10.1007/s100510050914}.

\bibitem[{\citenamefont{Kandpal et~al.}(2009)\citenamefont{Kandpal, Opahle,
  Zhang, Jeschke, and Valent\'\i}}]{Kandpal:2009}
\bibinfo{author}{\bibfnamefont{H.~C.} \bibnamefont{Kandpal}},
  \bibinfo{author}{\bibfnamefont{I.}~\bibnamefont{Opahle}},
  \bibinfo{author}{\bibfnamefont{Y.-Z.} \bibnamefont{Zhang}},
  \bibinfo{author}{\bibfnamefont{H.~O.} \bibnamefont{Jeschke}},
  \bibnamefont{and}
  \bibinfo{author}{\bibfnamefont{R.}~\bibnamefont{Valent\'\i}},
  \bibinfo{journal}{Phys. Rev. Lett.} \textbf{\bibinfo{volume}{103}},
  \bibinfo{pages}{067004} (\bibinfo{year}{2009}),
  \urlprefix\url{http://link.aps.org/doi/10.1103/PhysRevLett.103.067004}.

\bibitem[{\citenamefont{Chou and Nelson}(1996)}]{Nelson:1996}
\bibinfo{author}{\bibfnamefont{T.}~\bibnamefont{Chou}} \bibnamefont{and}
  \bibinfo{author}{\bibfnamefont{D.~R.} \bibnamefont{Nelson}},
  \bibinfo{journal}{Phys. Rev. E} \textbf{\bibinfo{volume}{53}},
  \bibinfo{pages}{2560} (\bibinfo{year}{1996}).

\bibitem[{\citenamefont{Brout}(1960)}]{Brout:1960}
\bibinfo{author}{\bibfnamefont{R.}~\bibnamefont{Brout}},
  \bibinfo{journal}{Phys. Rev.} \textbf{\bibinfo{volume}{118}},
  \bibinfo{pages}{1009} (\bibinfo{year}{1960}),
  \urlprefix\url{http://link.aps.org/doi/10.1103/PhysRev.118.1009}.

\bibitem[{\citenamefont{et~al.}(2007)}]{ALPS}
\bibinfo{author}{\bibfnamefont{A.~A.} \bibnamefont{et~al.}},
  \bibinfo{journal}{J. Magn. Magn. Mater.} \textbf{\bibinfo{volume}{310}},
  \bibinfo{pages}{1187} (\bibinfo{year}{2007}).

\end{thebibliography}

\end{document}